\documentclass{emulateapj}
\usepackage{amssymb}
\usepackage{graphicx}
\usepackage{grffile}
\usepackage{epsfig}
\usepackage{epstopdf}
\usepackage{url}

\newcommand{\msun}{\ensuremath{\rm M_\odot}}

\newcommand{\msunyr}{\ensuremath{\rm M_{\odot}\,{\rm yr}^{-1}}}

\newcommand{\Ha}{\ensuremath{\rm H\alpha}}
\newcommand{\Hb}{\ensuremath{\rm H\beta}}

\newcommand{\lya}{\ensuremath{\rm Ly\alpha}}
\newcommand{\wlya}{$W_{\rm Ly\alpha}$}

\newcommand{\NII}{[\ion{N}{2}]}
\newcommand{\kms}{km\,s\ensuremath{^{-1}}}
\newcommand{\ztwo}{\ensuremath{z\sim2}}

\newcommand{\OII}{[\ion{O}{2}]}
\newcommand{\OIII}{[\ion{O}{3}]}

\newcommand{\HII}{\ion{H}{2}}

\newcommand{\pasa}{PASA}

\begin{document}

\title{A High Fraction of \lya-Emitters Among Galaxies\\ with Extreme Emission Line Ratios at $z\sim2$\altaffilmark{*}}
\author{Dawn K. Erb\altaffilmark{1}, Max Pettini\altaffilmark{2}, Charles C. Steidel\altaffilmark{3},  Allison L. Strom\altaffilmark{3},\\ Gwen C. Rudie\altaffilmark{4}, Ryan F. Trainor\altaffilmark{5,6},   Alice E. Shapley\altaffilmark{7}, Naveen A. Reddy\altaffilmark{8,9}}

\slugcomment{DRAFT: \today}

\shorttitle{\lya\ Emission from Extreme Galaxies at $z\sim2$}
\shortauthors{Erb et al.}

\altaffiltext{*}{Based on data obtained at the W. M. Keck Observatory, which is operated as a scientific partnership among the California Institute of Technology, the University of California, and the National Aeronautics and Space Administration, and was made possible by the generous financial support of the W. M. Keck Foundation.}

\altaffiltext{1}{Center for Gravitation, Cosmology and Astrophysics, Department of Physics, University of Wisconsin Milwaukee, 3135 N. Maryland Avenue, Milwaukee, WI 53211, USA; \url{erbd@uwm.edu}}
\altaffiltext{2}{Institute of Astronomy, Madingley Road, Cambridge CB3 0HA, UK}
\altaffiltext{3}{Cahill Center for Astrophysics, California Institute of Technology, 1216 E. California Blvd., MS 249-17, Pasadena, CA 91125, USA}
\altaffiltext{4}{Carnegie Observatories, 813 Santa Barbara Street, Pasadena, CA 91101, USA}
\altaffiltext{5}{Department of Astronomy, University of California, Berkeley, 501 Campbell Hall, Berkeley, CA 94720, USA}
\altaffiltext{6}{Miller Fellow}
\altaffiltext{7}{University of California, Los Angeles, Department of Physics and Astronomy, 430 Portola Plaza, Los Angeles, CA 90095, USA}
\altaffiltext{8}{Department of Physics and Astronomy, University of California, Riverside, 900 University Avenue, Riverside, CA 92521, USA}
\altaffiltext{9}{Alfred P.\ Sloan Fellow}

\begin{abstract}

Star-forming galaxies form a sequence in the \OIII\ $\lambda$5007/\Hb\ vs.\ \NII\ $\lambda$6584/\Ha\ diagnostic diagram, with low metallicity, highly ionized galaxies falling in the upper left corner. Drawing from a large sample of UV-selected star-forming galaxies at \ztwo\ with rest-frame optical nebular emission line measurements from Keck-MOSFIRE, we select the extreme $\sim5$\% of the galaxies lying in this upper left corner, requiring log(\NII/\Ha)~$\leq-1.1$ and log(\OIII/\Hb)~$\geq0.75$.  These cuts identify galaxies with $12+\log(\rm{O/H}) \lesssim 8.0$, when oxygen abundances are measured via the O3N2 diagnostic. We study the \lya\ properties of the resulting sample of 14 galaxies. The mean (median) rest-frame \lya\ equivalent width is 39 (36) \AA, and 11 of the 14 objects (79\%) are \lya-emitters (LAEs) with \wlya~$> 20$ \AA. We compare the equivalent width distribution of a sample of 522 UV-selected galaxies at  $2.0<z<2.6$ identified without regard to their optical line ratios; this sample has mean (median) \lya\ equivalent width $-1$ ($-4$) \AA, and only 9\% of these galaxies qualify as LAEs. The extreme galaxies typically have lower attenuation at \lya\ than those in the comparison sample, and have $\sim50$\% lower median oxygen abundances. Both factors are likely to facilitate the escape of \lya: in less dusty galaxies \lya\ photons are less likely to be absorbed during multiple scatterings, while the harder ionizing spectrum and higher ionization parameter associated with strong, low metallicity star formation may reduce the covering fraction or column density of neutral hydrogen, further easing \lya\ escape. The use of nebular emission line ratios may prove useful in the identification of galaxies with low opacity to \lya\ photons across a range of redshifts. 

\end{abstract}

\keywords{galaxies: evolution---galaxies: formation---galaxies: high-redshift}

\section{Introduction}
\label{sec:intro}

Low metallicity galaxies may be the primary source of the radiation that reionized the universe.  It is likely that many faint and probably low metallicity galaxies were required to supply a sufficient number of photons \citep{kf12, rfs+13,befd15}, and ionizing radiation has been detected from galaxies with rest-frame optical line ratios indicative of low metallicity and high ionization parameter \citep{ios+16,vdv+16}. Until the launch of the \textit{James Webb Space Telescope (JWST)} enables studies of galaxies in the reionization era, the most effective way to analyze low metallicity sources is through the detailed examination of galaxies at $z\lesssim3$, when most of the strong rest-frame optical emission lines can be observed from the ground.

As the strongest feature in the ultraviolet (UV) spectra of many galaxies, the \lya\ emission line is of particular interest:\ its strength depends on a galaxy's dust content and on the kinematics, column density, geometry and covering fraction of neutral hydrogen, all quantities that are also likely to influence the escape of Lyman continuum radiation (see \citealt{d14} for a recent review).  However, the relative importance of these properties in setting the strength of \lya\ emission is still uncertain. 

\begin{figure*}[htbp]4959
\plotone{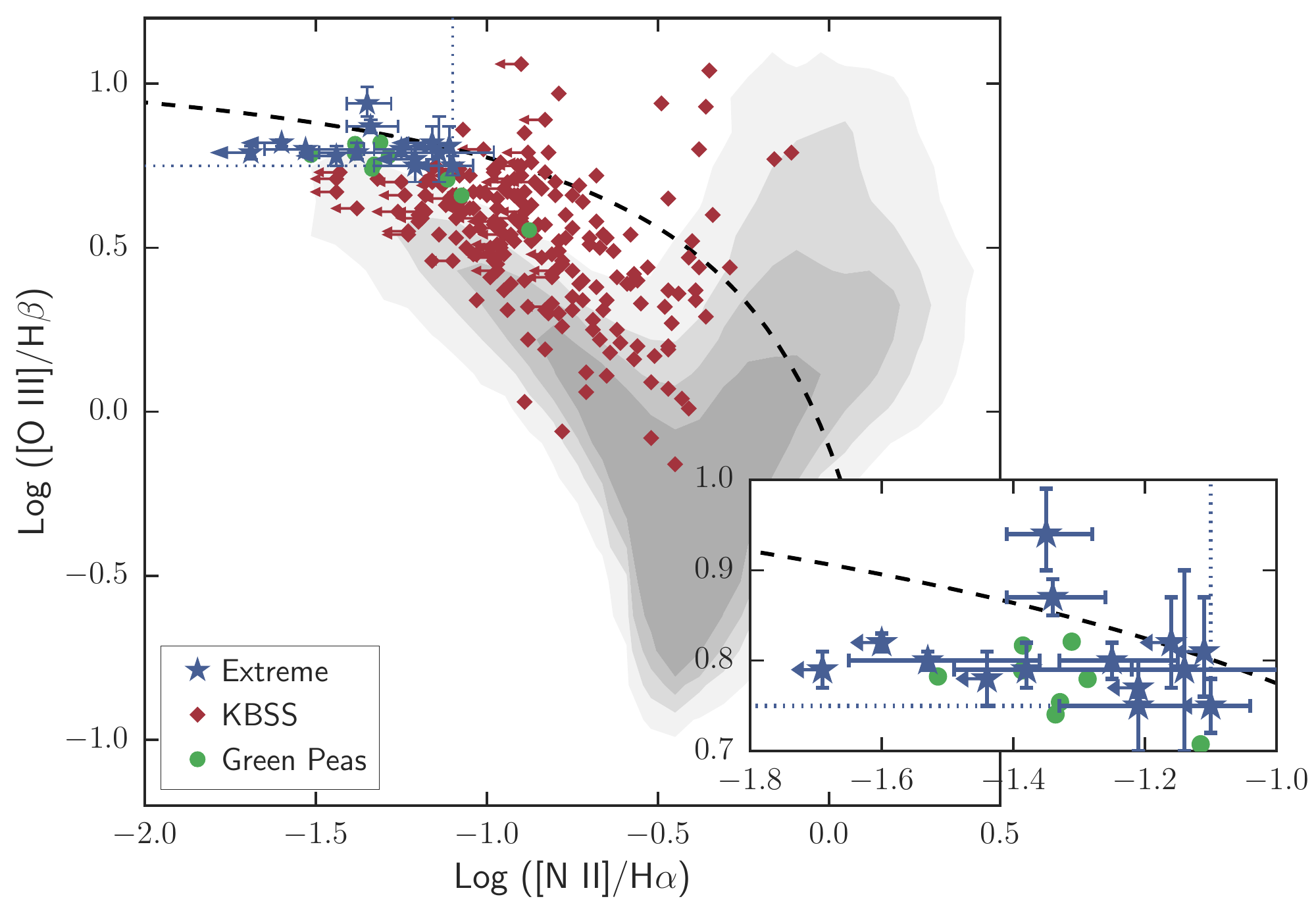}
\caption{The \OIII/\Hb\ vs.\ \NII/\Ha\ diagnostic diagram, in which star-forming galaxies fall on the left side and AGN in the upper right. The 14 galaxies in the extreme sample (blue stars) lie in the upper left corner, with log (\NII/\Ha)~$\leq-1.1$ and log (\OIII/\Hb)~$\geq0.75$ (dotted blue lines). We also show the KBSS parent sample of \citet{srs+14} with red diamonds and Green Pea galaxies from \citet{hsme15} with green circles. Grey contours show local galaxies from the Sloan Digital Sky Survey, and the black dashed line shows the maximum starburst line of \citet{kds+01}. The inset panel shows a closeup view of the extreme region for clarity.}
\label{fig:bpt}
\end{figure*}

Low metallicity galaxies at \ztwo\ that have been studied in detail typically have very strong \lya\ emission (rest-frame \lya\ equivalent width \wlya~$\gtrsim 50$ \AA; \citealt{eps+10,stark+14}), but galaxies with high \lya\ equivalent widths span a wide range in mass and can be old or dusty (e.g.\ \citealt{kse+10, hcg+14}).  Most \lya-emitters (LAEs, defined as having  \wlya~ $>20$ \AA\ and usually selected via their \lya\ emission) are fainter and lower in mass than galaxies selected via broad-band continuum colors \citep{gvg+06,gap+11,mrm+14}, and these mass and luminosity differences between samples complicate efforts to determine the primary factors influencing \lya\ emission. A useful technique is therefore the comparison of matched (in stellar mass or continuum magnitude) galaxy samples with and without \lya\ emission; using this method, \citet{kse+10} found that \lya-emitting galaxies at $z\sim3$ are less dusty but older than comparable galaxies without \lya\ emission, while \citet{hzb+16} recently found no statistically significant differences between \lya-selected and \OIII-selected galaxies of the same masses at \ztwo.
 
Because high redshift LAEs are faint, few have measurements of metallicity, ionization parameter, or other properties relying on the detection of weak nebular emission lines. Those measured tend to be at the bright end of the distribution, and generally have low metallicities, high ionization parameters, and little dust \citep{nos+13, sfg+14}. This appears to be true in the local universe as well:\ low metallicity galaxies with high equivalent width optical emission lines (the ``Green Peas,"  \citealt{igt11}) have strong \lya\ emission \citep{hsme15}, while \citet{cbh11} find that LAEs have lower metallicities, bluer colors, and higher \Ha\ equivalent widths than galaxies with similar UV continuum magnitudes but without \lya\ emission.

In this paper we examine the \lya\ properties of a sample of galaxies at \ztwo\ selected to have extreme ratios of rest-frame optical emission lines. We identify galaxies lying in the upper left corner of the  \OIII\ $\lambda$5007/\Hb\ vs.\ \NII\ $\lambda$6584/\Ha\ diagnostic diagram (the ``BPT" diagram, \citealt{bpt81}), which represents the low metallicity, high ionization end of the star-forming sequence. We then use the remainder of the parent sample as a comparison sample for the extreme objects; relative to the extreme sample, this comparison sample has less extreme line ratios but approximately similar masses and continuum magnitudes.

We describe our data and sample selection in Section \ref{sec:data}. In Section \ref{sec:optical} we discuss the rest-frame optical emission line properties of the galaxies in the sample, and we quantify their \lya\ emission in Section \ref{sec:lya}. We summarize and discuss our results in Section \ref{sec:summary}. We assume a \citet{c03} initial mass function and a cosmology with $H_0=70$ \kms\ Mpc$^{-1}$, $\Omega_{\rm m}=0.3$, and $\Omega_{\Lambda}=0.7$ throughout. 

\section{Data and Sample Selection}
\label{sec:data}

The galaxies discussed in this paper are drawn from the Keck Baryonic Structure Survey (KBSS; \citealt{ssp+04,srs+14}), for which objects are initially selected via rest-frame UV color criteria designed to identify star-forming galaxies at redshifts $2\lesssim z \lesssim 2.6$. The parent sample for this work consists of 315 galaxies in the redshift range $1.95\lesssim z \lesssim 2.65$ for which measurements (or upper limits in the case of \NII\ $\lambda$6584) of \Ha, \NII\ $\lambda$6584, \OIII$ \lambda$5007 and \Hb\ have been obtained with the Multi-Object Spectrometer for InfraRed Exploration (MOSFIRE; \citealt{mosfire,mosfire2}) on the Keck I telescope. An earlier version of this parent sample, with 219 galaxies, was described in detail by \citet{srs+14}, and we refer the reader to that paper for details of the MOSFIRE observations and data reduction.

For this work, we select the extreme $\sim5$\% of the KBSS sample based on position in the \OIII/\Hb\ vs.\ \NII/\Ha\ (BPT) diagnostic diagram. Low metallicity, highly ionized galaxies lie in the upper left corner, and the most extreme $\sim5$\% have both log(\NII/\Ha)~$\leq-1.1$ and log(\OIII/\Hb)~$\geq0.75$ (see Figure \ref{fig:bpt}). These thresholds are motivated by the general position of the extreme galaxy Q2343-BX418 \citep{eps+10} in the BPT diagram, as we wish to identify similar objects.

Fourteen galaxies with redshifts $2.17 < z < 2.47$ and mean (median) redshift $z=2.30$ (2.30) fall in this region of the diagram, including seven objects with 3$\sigma$ upper limits on \NII\ emission such that log (\NII/\Ha)~$\leq-1.1$. We refer to these 14 galaxies as the extreme sample, and the galaxies and their rest-frame optical emission line properties are listed in Table \ref{tab:neblines}.

Most of the rest-frame UV spectroscopic observations of the 14 extreme galaxies were obtained prior to the MOSFIRE observations, using the the Low Resolution Imaging Spectrometer (LRIS, \citealt{occ+95,ssp+04}) on the Keck I telescope. LRIS observations and data reduction are described by \citet{ssp+04}. Eight of the 14 galaxies (Q0100-BX239, Q0142-BX165, Q0207-BX74, Q0207-BX87, Q0207-BX144, Q0821-BX102, Q0821-BX221, and Q1700-BX711) were observed with the LRIS-B 400-line grism, offering spectral resolution of $\sim500$ \kms, while the remaining 6 (Q1217-BX164, Q1700-BX553, Q2206-BM64, Q2343-BX418, Q2343-BX460, and Q2343-BX660) were observed with the 600-line grism, providing $\sim300$ \kms\ resolution. 
Most of the objects were observed with the standard KBSS integration time of $3\times1800$ seconds, but Q1700-BX553, Q2343-BX418, Q2343-BX460, and Q2343-BX660 were placed on masks that received between 5 and 11 hours of integration. Because of the varying integration times, the rest-frame UV spectra of the extreme subsample vary widely in S/N; for this reason, we focus here on the \lya\ properties of the galaxies, the only feature that can reliably be measured in the full sample. A more detailed analysis of the extreme objects with high quality rest-frame UV spectra will be presented separately.

\section{Inferences from Rest-Frame Optical Emission Lines}
\label{sec:optical}

As shown in Figure \ref{fig:bpt}, the extreme sample lies in the upper left corner of the BPT diagram. The 14 extreme galaxies have mean (median) stellar mass $6.2 \times 10^9$ ($4.0 \times 10^9$) \msun, and mean (median) dust-corrected \Ha\ star formation rate 36.6 (27.8) \msunyr\ (see Table \ref{tab:neblines}; star formation rates are calculated assuming the \citet{k98} relationship between SFR and \Ha\ luminosity). Their median rest-frame UV absolute magnitude is $M_{\rm UV} = -20.66$, indicating that these are $L^*$ galaxies; \citet{rs09} find $M^* = -20.70$ at \ztwo, and all of the extreme objects are within $\sim1$ mag of this value.

We show several other galaxy samples in Figure \ref{fig:bpt} for comparison. The \ztwo\ KBSS parent sample of \citet{srs+14} is shown in red. With the extreme galaxies excluded, the KBSS galaxies are similar to the extreme sample in mass and SFR, with a somewhat higher mean (median) mass of $2.0 \times 10^{10}$ ($9.7 \times 10^9$) \msun\ and slightly lower mean (median) SFR of 36.1 (23.6) \msunyr. Stellar masses are computed by modeling the spectral energy distribution from the rest-frame UV to near-IR, with broadband magnitudes corrected for the contribution of nebular emission lines. We use the \citet{bc03} stellar population models with solar metallicity, the \citet{cab+00} extinction law, and assume a constant star formation history; detailed descriptions of the modeling procedure and discussions of the systematic uncertainties associated with these choices are given by \citet{sse+05} and \citet{rps+12}.

We also show ten local ``Green Pea" galaxies in green \citep{hsme15}, and contours showing the locus of Sloan Digital Sky Survey galaxies in grey. The Green Peas are drawn from the SDSS, with selection based on very high equivalent width rest-frame optical emission lines \citep{css+09,igt11}. The ten Green Peas shown are somewhat lower in mass and SFR than the extreme \ztwo\ sample, with median stellar mass and SFR $1.0 \times 10^9$ \msun\ and 15.2 \msunyr\ respectively. The extreme sample lies in a region of the BPT diagram in which there are relatively few local galaxies; however, six of the ten Green Peas fall within the extreme selection box, and only one is $>0.1$ dex from the extreme cuts in either \NII/\Ha\ or \OIII/\Hb. We mark the Green Peas that meet the extreme BPT selection criteria with open circles in later figures.

We estimate the metallicity of the extreme galaxies using the O3N2 indicator proposed by \citet{pp04}, which \citet{srs+14} show to be in reasonable agreement with metallicities derived from the $T_e$ method (using the \ion{O}{3}] $\lambda\lambda 1661$,1666 auroral lines)  for three of the galaxies in the extreme sample. This calibration has the advantage of using only closely spaced pairs of emission lines, and is therefore insensitive to uncertainties in band-to-band calibration or reddening; however, because we report 3$\sigma$ upper limits on \NII\ for half of the sample, we determine upper limits on metallicity for 7/14 galaxies.

If we count upper limits as detections, the extreme sample occupies a narrow range in metallicity, with $12+\log$(O/H) ranging from 7.85 to 8.12 and mean (median) oxygen abundance 8.03 (8.06). For comparison, the \ztwo\ sample of \citet{srs+14}, not including the 14 extreme galaxies, has mean (median) O3N2 metallicity 8.29 (8.27), again counting \NII\ limits as detections. 

The metallicities of the extreme \ztwo\ sample are very similar to the $T_e$-based metallicities of the Green Pea comparison sample, which range from 7.86 to 8.17 and have an identical mean of 8.03 \citep{igt11,hsme15}.  We use this Green Pea sample to assess the reliability of the O3N2 indicator by comparing the O3N2 metallicities of the Green Peas with their direct abundances measured from the \OIII\ $\lambda 4363$ line. We find that the O3N2 metallicities are lower than the $T_e$ metallicities by 0.06 dex on average, seven of the ten Green Peas agree within 0.1 dex, and only one object disagrees by $>0.2$ dex; this object is also the farthest from the extreme selection box in the BPT diagram by a considerable margin. We conclude that for galaxies in this range of parameter space, the O3N2 diagnostic appears to be a reliable indicator of the gas phase oxygen abundance.

The similar metallicities of the extreme galaxies are a consequence of the selection criteria, since each point on the \OIII/\Hb\ vs.\ \NII/\Ha\ diagram directly corresponds to a metallicity determined with the O3N2 indicator. In this sense, the extreme sample is metallicity-selected. The oxygen abundances of the extreme sample are listed in Table \ref{tab:neblines}. 

\begin{figure}[htbp]
\centerline{\epsfig{angle=00,width=\hsize,file=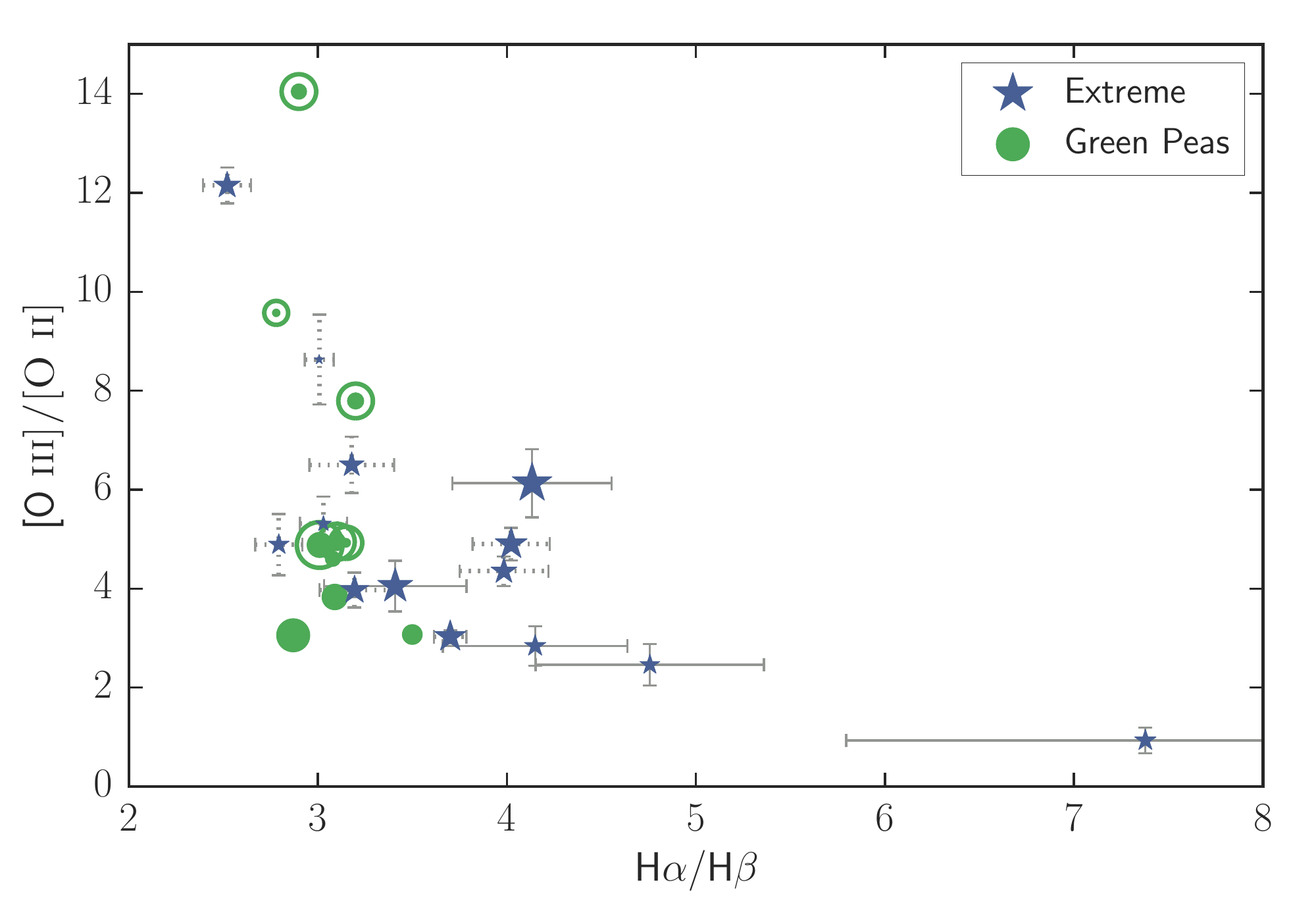}}
\caption{The extinction-corrected \OIII\  $\lambda\lambda 4959, 5007$/\OII\ $\lambda\lambda 3727, 3729$ ratio vs. \Ha/\Hb\ for the extreme and Green Pea samples. 
We identify the extreme galaxies with Balmer decrements measured with $>10\sigma$ significance with dashed error bars; the five objects with \Ha/\Hb\ ratios measured with less than 10$\sigma$ significance are marked with solid error bars. Green Peas which meet the extreme BPT selection criteria are circled. The sizes of the symbols increase with increasing \lya\ equivalent width (see Section \ref{sec:lya}).}
\label{fig:bd_o3o2}
\end{figure}

We also use the nebular line measurements to assess the extinction and ionization properties of the extreme sample, from the \Ha/\Hb\ and \OIII/\OII\  ratios respectively. Both of these ratios employ measurements made in two different bands (the $H$ and $J$ bands for \OIII\ and \OII, and the $K$ and $H$ bands for \Ha\ and \Hb). Band-to-band corrections for slit losses (based on stars placed on each mask, modeling of the line profiles, comparison of the spectroscopic continuum to broadband photometry, and multiple observations of the same objects) have been applied, but these ratios are subject to larger systematic uncertainties than the ratios of closely spaced lines such as \OIII/\Hb\ and \NII/\Ha. The median correction is a factor of 2 in all three bands, with a median uncertainty of 6\%; all $H$- and $K$-band corrections have $<20$\% uncertainties, as do all but two of the $J$-band corrections (Q0821-BX102 and Q1700-BX553 have 27\% and 28\% uncertainties respectively). The slit-loss corrections are discussed in full elsewhere (\citealt{srs+14}; Strom et al.\ in preparation). \Hb\ fluxes are also corrected for underlying absorption by a factor of 1.06.

The  \Ha/\Hb\ ratios of the extreme sample and their statistical errors are listed in Table \ref{tab:neblines}.  In contrast to the Green Peas, which lie close to the Case B value\footnote{In this work we adopt an intrinsic ratio $I(\Ha)/I(\Hb) = 2.89$, which Steidel et al.\ (2016, in prep) find is predicted by photoionization models that best reproduce the observed line ratios of a subsample of KBSS galaxies with high quality rest-frame UV and optical spectra.} with mean and standard deviation $\langle$\Ha/\Hb$\rangle = 3.07 \pm 0.19$, the extreme galaxies show considerable variation in \Ha/\Hb, with half of the sample having \Ha/\Hb~$>3.5$. We find mean and standard deviation $\langle$\Ha/\Hb$\rangle = 3.80 \pm 1.16$, with values ranging from below Case B to a maximum of 7.38. However, we note that the objects with the highest \Ha/\Hb\ ratios have the largest uncertainties. The reliability of the Balmer decrement measurements will be discussed in detail in a forthcoming paper (Strom et al.\ 2016, in prep); here, we note that Strom et al.\ find that extinction corrections for galaxies with measurements of the Balmer decrement with less than 10$\sigma$ significance are less reliable than those with higher S/N.\footnote{This determination is based on the finding that galaxies with extinction corrections based on a $>10 \sigma$ measurement of the Balmer decrement form a tight sequence in the dust-corrected O$_{32}$-R$_{23}$ plane, while many objects with less significant \Ha/\Hb\ measurements are outliers, likely due to an error in the extinction correction.} We therefore adopt this threshold in considering whether our \Ha/\Hb\ measurements are robust. Nine of the 14 extreme galaxies have $>10\sigma$ Balmer decrement measurements, and we find lower extinction in this more robust subsample, with mean (median) \Ha/\Hb = 3.27 (3.18) and standard deviation 0.49. These results suggest that the extreme \ztwo\ galaxies have a somewhat wider range in nebular line extinction than the Green Peas, but a larger sample with very high S/N Balmer decrement measurements is required in order to confirm and quantify this result.

\begin{figure*}[htbp]
\centerline{\epsfig{angle=00,width=\hsize,file=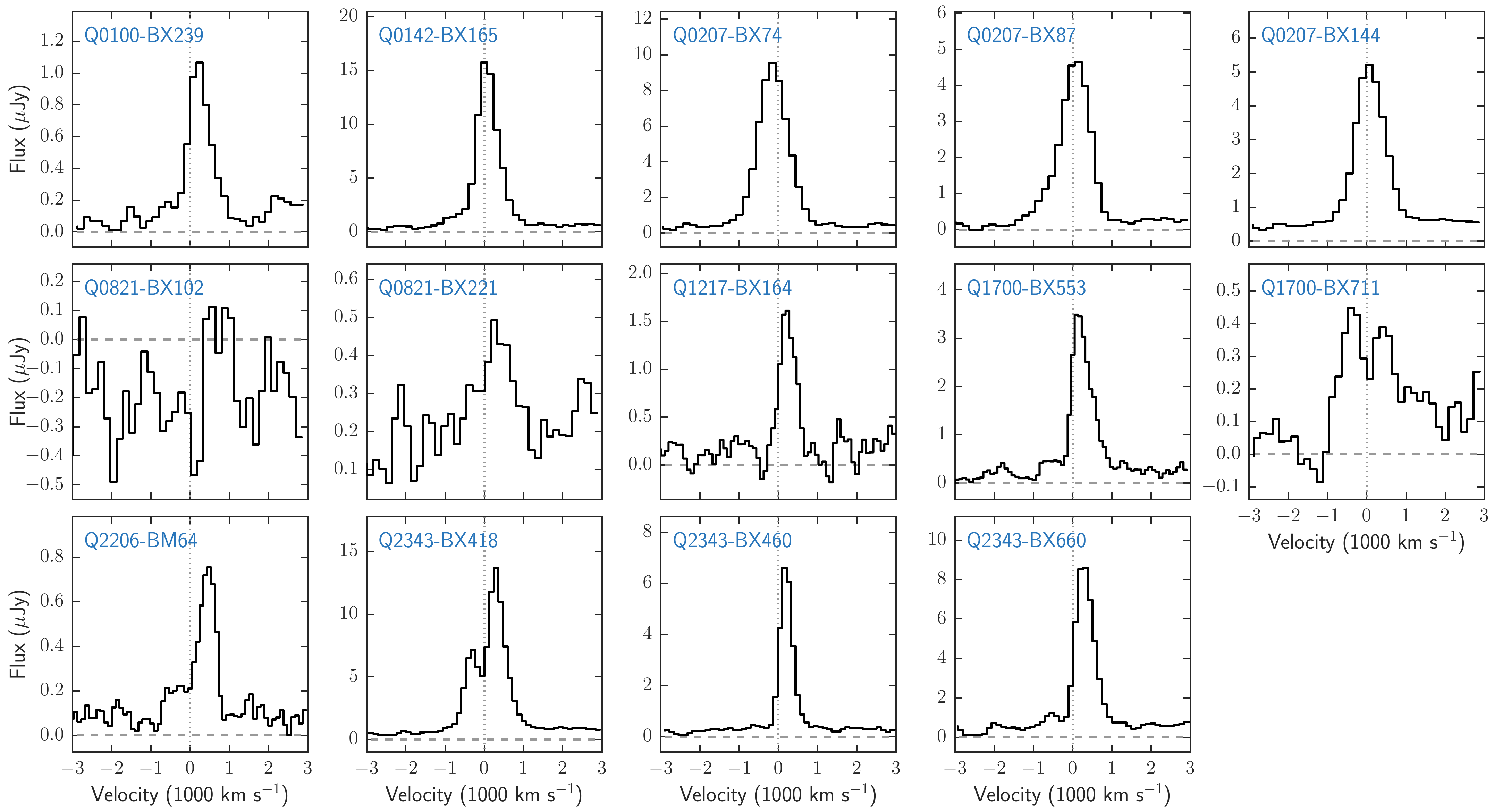}}
\caption{\lya\ profiles of the 14 extreme galaxies. Vertical dotted lines show the systemic redshift measured from nebular emission lines. Most of the extreme galaxies show strong \lya\ emission.}
\label{fig:lyaprofiles}
\end{figure*}

We estimate the ionization states of the extreme and Green Pea samples via the \OIII/\OII\ ratio, which increases with increasing ionization parameter with some additional dependence on metallicity (e.g.\ \citealt{kd02}). We measure $\mbox{O}_{32} \equiv \mbox{F}($\OIII$ \lambda\lambda 4959, 5007)/\mbox{F}($\OII$ \lambda\lambda 3727, 3729$). The observed \OIII/\OII\ ratios of the extreme galaxies are listed in Table \ref{tab:neblines}; values range between 2.4 and 12.2, with a median value of 5.2. We then correct the \OIII/\OII\ ratios for extinction, using the Balmer decrement measurements and the \citet{ccm89} extinction law with $R_V=3.1$. The median extinction-corrected \OIII/\OII\ ratio is 4.6, while the extreme galaxies with $>10\sigma$ Balmer decrements have median 4.9. These values are significantly higher than those of more typical star-forming galaxies at the same redshift; for the parent KBSS sample, Strom et al.\ (in prep) find median $\mbox{O}_{32} = 1.7$ for galaxies with $>10\sigma$ measurements of the Balmer decrement (and median $\mbox{O}_{32} = 1.4$ for the full sample), while \citet{ssk+15}
use a sample of $z\sim2.3$ galaxies from the MOSDEF survey \citep{ksr+15} and find median $\mbox{O}_{32} = 1.3$, with a 68\% upper bound of 2.3.  As expected, the extreme line ratio criteria identify highly ionized galaxies at \ztwo. 

The extreme galaxies have \OIII/\OII\ ratios similar to the Green Peas:\ the ten galaxies studied by \citet{hsme15} have median $\mbox{O}_{32} = 4.9$, while the subset that meet the extreme BPT selection criteria have median $\mbox{O}_{32} = 6.4$. The Green Peas are also much more highly ionized than typical galaxies in the local universe; \citet{ssk+15} report median $\mbox{O}_{32} = 0.3$ for their $0.04<z<0.1$ SDSS comparison sample.

The extinction-corrected \OIII/\OII\ ratios are plotted against \Ha/\Hb\ for both the extreme sample and the Green Peas in Figure \ref{fig:bd_o3o2}. The Green Peas and most of the extreme galaxies with $>10\sigma$ Balmer decrement measurements occupy a similar region in this diagram, but the extreme sample shows a lower S/N tail of objects with high \Ha/\Hb\ and low \OIII/\OII. Because the extinction corrections for these objects are more uncertain, it is not yet clear whether the low \OIII/\OII\ ratios are due to an overestimate of the extinction. It is also notable that the extreme BPT selection criteria identify the Green Peas with the highest values of $\mbox{O}_{32}$. 

We conclude from these nebular line measurements that the extreme selection reliably identifies galaxies with low ($\sim20$\% of solar) metallicities and ionization states consistently higher than those of more typical \ztwo\ galaxies. 
The extreme galaxies also have less dust attenuation at \lya\ than more typical \ztwo\ galaxies; however, there is considerable range in the \Ha/\Hb\ ratio, and a more robust measurement of the variation in extinction requires a larger sample with high S/N measurements of the Balmer decrement. We discuss the nebular emission line properties of the extreme sample further in the context of the \lya\ measurements described below.

\begin{figure*}[htbp]
\plotone{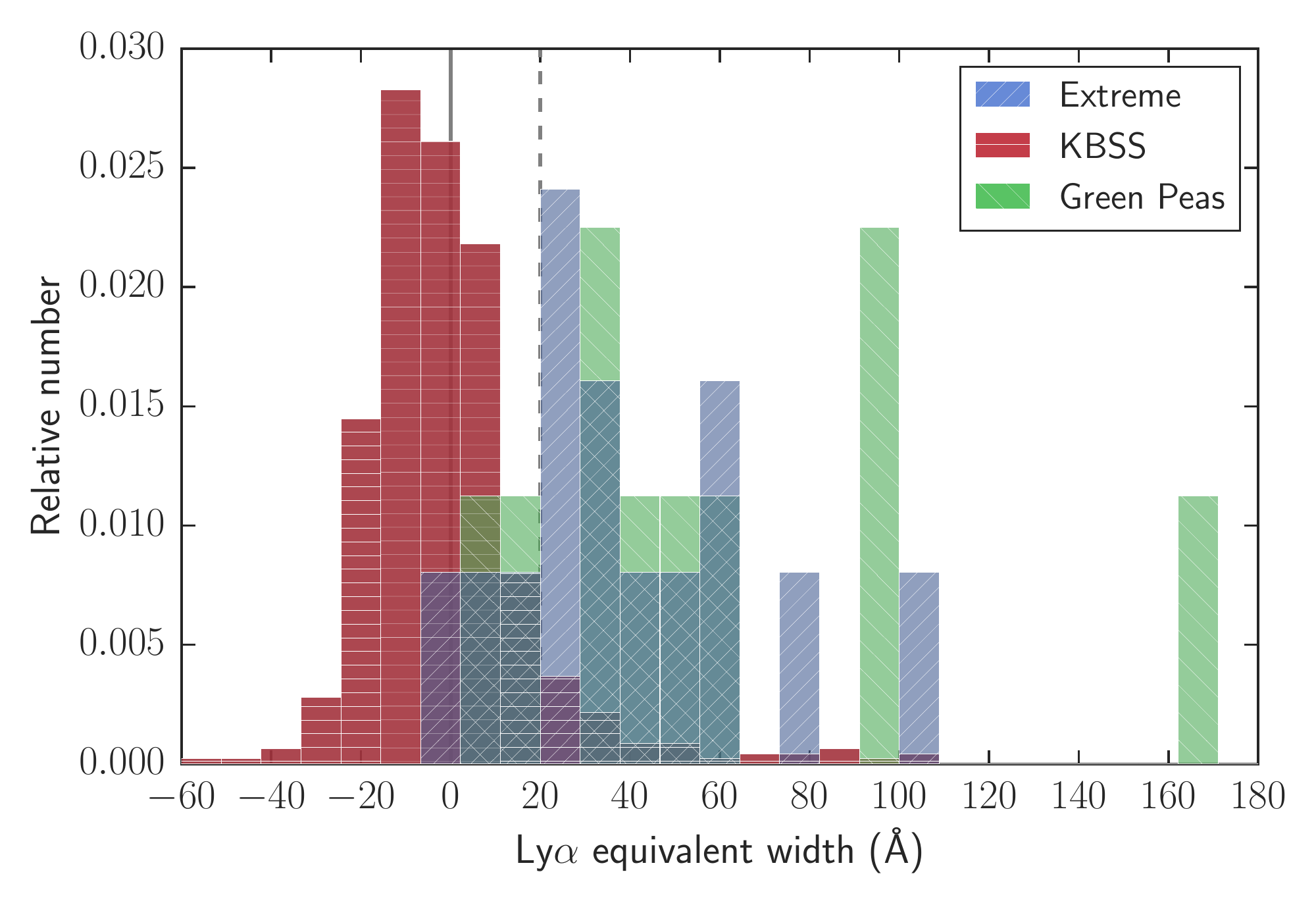}
\caption{\lya\ rest-frame equivalent width distributions for the extreme sample (blue), 522 $2.0\leq z \leq 2.6$ galaxies from the KBSS parent sample (red), and Green Pea galaxies (green, \citealt{hsme15}). Histograms are normalized to an integral of unity. The extreme and Green Pea samples show similar \wlya\ distributions, while the KBSS comparison sample typically shows much weaker \lya\ emission.}
\label{fig:lyahists}
\end{figure*}

\section{\lya\ Properties}
\label{sec:lya}

The \lya\ profiles of the 14 extreme galaxies are shown in Figure \ref{fig:lyaprofiles}, from which it is clear that most of the galaxies in the sample show \lya\ in emission. The exceptions are the two galaxies Q0821-BX102 and Q0821-BX221, which have weak or no emission. BX221 has the lowest SFR in the sample, though only slightly; otherwise these objects appear similar to the rest of the extreme sample.  Among the remaining 12 galaxies with strong emission, the lines show a variety of spectral morphologies:\ most are strong and relatively symmetric, while a few show multiple components. This diversity is consistent with previous observations of galaxies with strong \lya\ emission (e.g.\ \citealt{ksk+12,cbh+13,est+14,tssr15}). 

At the moderately low spectral resolution of our observations, Q2343-BX418 (measured with $\sim300$ \kms\  resolution) and Q1700-BX711 (measured with $\sim500$ \kms\ resolution) show double peaks:\ the red peak is significantly stronger in BX418, while BX711 has a slightly stronger blue peak and may have multiple components of redshifted emission. Q2206-BM64 also shows emission blueward of zero velocity. These objects are not outliers among any of the properties we measure in this work, indicating that the \lya\ emission profile depends in part on factors which are not reflected by the nebular line ratios. We will explore this issue further with a detailed analysis of the highest S/N rest-frame UV spectra in the extreme sample, focusing on the interstellar absorption lines as well as on \lya\ (Erb et al.\ in prep).

We measure \lya\ fluxes and equivalent widths following the procedure described by \citet{kse+10}, in which each line profile is classified as emission, absorption, combination or noise, the continuum levels on the red and blue sides of the line are determined separately, and the equivalent width is calculated to be the flux in the line divided by the red continuum level. Uncertainties are determined via Monte Carlo simulations in which each spectrum is perturbed 500 times by a random value drawn from a Gaussian distribution with width corresponding to the uncertainty at each pixel. 
We repeat the flux and equivalent width measurements for each perturbed spectrum, and the medians and standard deviations of these measurements are taken as the flux, equivalent width, and their uncertainties. For the extreme sample, all galaxies are classified as emission, with the exception of Q0821-BX102 which is classified as noise. 

The resulting distribution of \lya\ rest-frame equivalent widths is shown in Figure \ref{fig:lyahists}, and \lya\ measurements are given in Table \ref{tab:lya}. The mean (median) \lya\ equivalent width of the extreme sample is 39.3 (35.6) \AA, and 11 of the 14 galaxies (79\%) have \wlya~$>20$ \AA, the nominal threshold to be classified as an LAE, while an additional object has \wlya~$=18.0 \pm 2.0$ \AA.

We show two other galaxy samples in Figure \ref{fig:lyahists} for comparison. The dark red histogram shows the \lya\ equivalent width distribution of 522 UV-selected KBSS galaxies at $2.0\leq z \leq 2.6$ with mean redshift $\langle z \rangle = 2.28 \pm 0.16$. These galaxies were selected without regard to their nebular line ratios, and equivalent widths were measured using the same procedure as for the extreme sample, with \lya\ spectral morphologies independently classified by three of the authors (DKE, NAR, AES). The mean (median) \lya\ equivalent width of this comparison sample is $-1.0$ ($-4.4$) \AA, and 9\% of the galaxies in the sample qualify as LAEs with  \wlya~$>20$ \AA. A two sample K-S test indicates that the extreme and KBSS comparison samples are statistically different:\ the probability that the two samples are drawn from the same equivalent width distribution is $P=8.0\times10^{-8}$  (5.4$\sigma$ significance).

We also compare the equivalent widths of the Green Pea galaxies (\citealt{hsme15}; green histogram). In contrast to the KBSS sample, the extreme \ztwo\ galaxies and the Green Peas have similar equivalent width distributions; the mean (median) \wlya\ of the Green Peas is 59.6 (44.0) \AA, and a K-S test finds that the probability that the extreme \ztwo\ galaxies and the Green Peas are drawn from the same distribution of equivalent widths is $P=0.59$ (0.54$\sigma$ significance).

Finally, we measure the offset of the centroid of the \lya\ emission line\footnote{Velocity offsets are measured from the flux-weighted centroid of the total \lya\ emission in all cases. A forthcoming analysis of objects with higher S/N spectra will address the velocities of individual components of the \lya\ emission.} from the systemic velocity, finding a mean (median) velocity offset of 186 (208) \kms. This offset is typical of LAEs at \ztwo, which generally have smaller velocity offsets than continuum-selected galaxies at the same redshifts \citep{hos+13,son+14,est+14,tssr15}; similarly, continuum-selected $z\sim3$ galaxies with strong \lya\ emission show smaller velocity differences between the interstellar absorption lines and \lya\ than galaxies with lower \lya\ equivalent widths \citep{ssp+03}.  

These comparisons clearly show that galaxy selection via nebular emission line properties can yield a high fraction of objects with strong \lya\ emission. Previous studies of small samples of relatively bright LAEs have indicated that galaxies with strong \lya\ emission tend to have low metallicities \citep{nos+13, sfg+14}; we now see that the reverse is also true, reinforcing the connection between the physical conditions in \HII\ regions and the escape of \lya\ photons.

\subsection{Ly$\alpha$ Emission and Extinction}
\label{sec:extinct}

Many studies have shown that galaxies with strong \lya\ emission have relatively low extinction, as measured either from their bluer UV slopes or lower \Ha/\Hb\ ratios (e.g.\ \citealt{ssp+03, gvg+06, vgd+09, kse+10, pgs+10, gap+11, aks+14, mrm+14, hsme15}, cf.\ \citealt{hzb+16}). The extreme \ztwo\ sample shows a range in the Balmer decrement, suggesting some range in internal extinction. 

As discussed in Section \ref{sec:optical}  above, we use the \citet{ccm89} extinction law to calculate $E(B-V)_{\rm neb}$ for the extreme sample. The 14 galaxies have median \Ha/\Hb~$ = 3.56$ with a standard deviation of 1.16, which results in median $E(B-V)_{\rm neb} = 0.20\pm0.23$. If we consider only the subsample with $>10\sigma$ measurements of the Balmer decrement, we find median \Ha/\Hb~$ = 3.18$ and $E(B-V)_{\rm neb} = 0.09$. For comparison, \citet{srs+14} report median \Ha/\Hb~$ = 3.89\pm0.65$ for the KBSS parent sample, for a corresponding color excess of $E(B-V)_{\rm neb} = 0.28 \pm 0.16$. These values indicate that the median extinction at the wavelength of \lya\ is lower in the extreme sample than in the KBSS comparison sample; it is 1.9 times lower when we compare the full extreme sample, and 4.3 times lower if only the $>10\sigma$ extreme sample is used.

Measurements of extinction from SED fitting (adopting the \citealt{cab+00} extinction law) also indicate lower attenuation at \lya\ in the extreme sample relative to the KBSS comparison sample, although the difference is smaller. The extreme objects have median $E(B-V)_{\rm cont} = 0.11$, while the comparison sample has median $E(B-V)_{\rm cont} = 0.17$. This difference corresponds to a factor of 1.6 lower attenuation at \lya\ in the extreme sample.

The Green Pea galaxies, which are similar to the extreme sample in their metallicities and \lya\ equivalent widths, have uniformly low extinction, with median \Ha/\Hb~$ = 3.07$. This is roughly comparable to the $>10\sigma$ extreme sample, which has 1.3 times higher median extinction at \lya; if the full extreme sample is used, we find that the median extinction at \lya\ is 2.9 times higher for the extreme galaxies than for the Green Peas. The \Ha/\Hb\ ratios for both the extreme sample and the Green Peas are shown in Figure \ref{fig:bd_o3o2}, plotted against the extinction-corrected \OIII/\OII\ ratio. The symbol sizes are scaled by the \lya\ equivalent width, from which it is apparent that the strength of \lya\ emission is uncorrelated with both the Balmer decrement and O$_{32}$. 

We conclude that the lower extinction of the extreme galaxies relative to the KBSS parent sample likely facilitates the escape of \lya\ photons, but some of the extreme objects may have both moderate to high extinction and strong \lya\ emission.  This combination is not surprising:\ although most galaxies with strong \lya\ emission have little dust, \citet{kse+10} and \citet{hcg+14} find LAEs with $E(B-V)$ up to 0.3 and 0.4 respectively. It is perhaps more unexpected for low metallicity objects with line ratios similar to the Green Peas to show significant extinction, and we await higher S/N measurements of the Balmer decrement to confirm whether this is indeed the case. 

\subsection{The Ly$\alpha$ Escape Fraction}

\begin{figure}[htbp]
\centerline{\epsfig{angle=00,width=\hsize,file=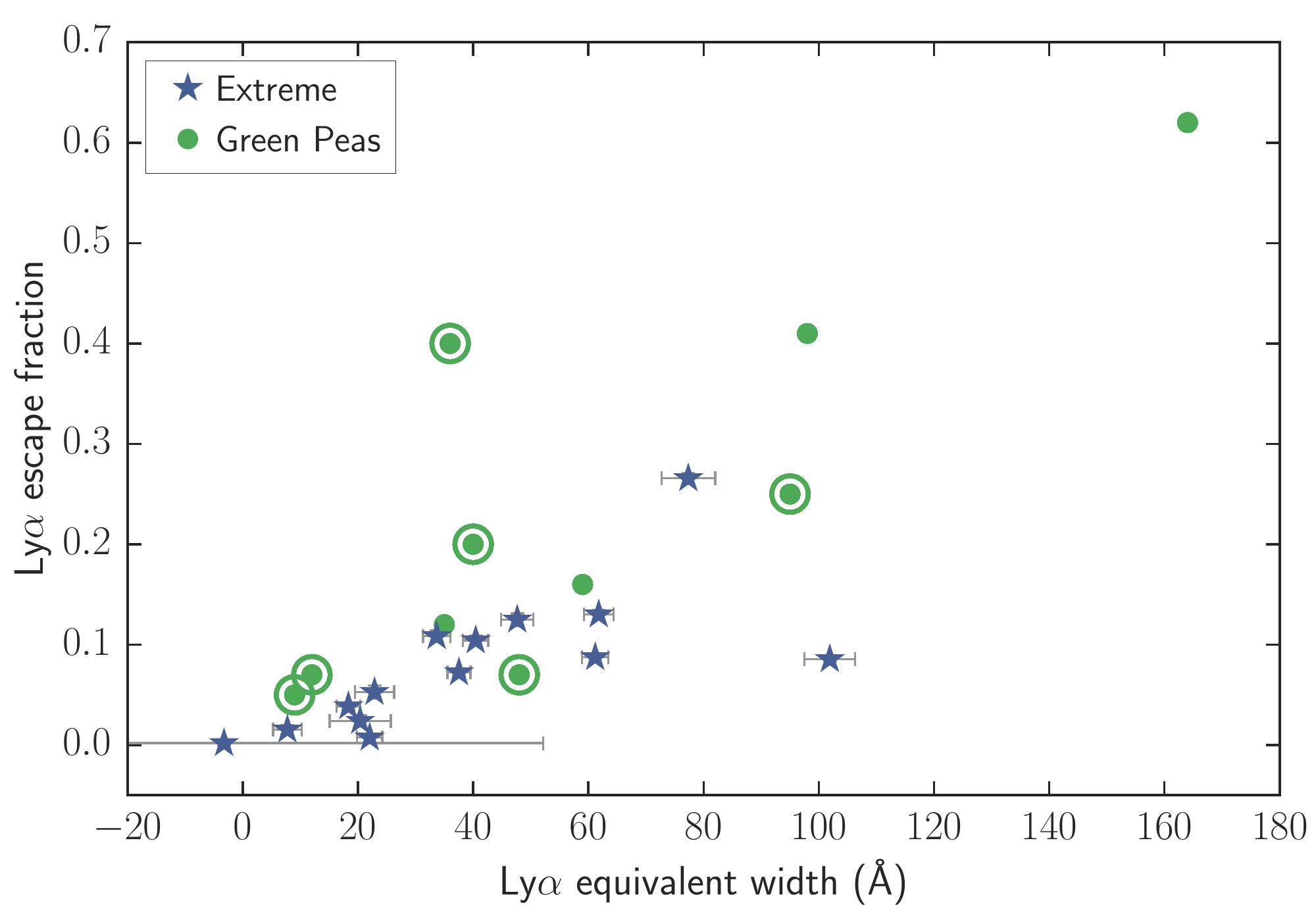}}
\caption{The \lya\ escape fraction vs.\ the \lya\ equivalent width for the extreme and Green Pea samples. For both samples, the escape fractions shown are not corrected for the fraction of \lya\ photons which escape the spectroscopic aperture; see discussion in text. Green Peas which meet the extreme selection criteria in the BPT diagram are circled.}
\label{fig:wlya_fesc}
\end{figure}

We estimate the fraction of \lya\ photons reaching the detector from the \lya/\Ha\ flux ratio, assuming an intrinsic ratio of \lya/\Ha~$=8.7$ \citep{hsme15,tssr15} and correcting the \Ha\ fluxes for extinction as described in Section \ref{sec:extinct} above. The UV spectra are corrected for slit losses by normalizing to the $G$-band magnitude; as discussed below, this does not account for additional losses in \lya\ due to possible differences in spatial distribution relative to the continuum. The \Ha\ fluxes are also corrected for slit losses, as we describe in Section \ref{sec:optical} above. We find that the escape fraction is typically low, with mean and median values of 0.08; the largest value in the sample is 0.27. These escape fractions are listed in Table \ref{tab:lya} and plotted vs.\ \lya\ equivalent width in Figure \ref{fig:wlya_fesc}.

For larger populations of star-forming galaxies at \ztwo, \citet{sbs+11} and \citet{hos+09} find \lya\ escape fractions of $\sim5$\%, while \citet{czg+14} find $f_{\rm esc}^{\lya} < 6$\%, only slightly lower than our mean value of 0.08. LAEs typically have higher escape fractions: \citet{hos+09} and \citet{sbs+11} find $f_{\rm esc}^{\lya} \sim 0.3$ for the LAEs in their samples, and \citet{tssr15} also estimate an escape fraction of $\sim30$\% for a sample of faint LAEs at $z\sim2.7$. However, in contrast to the current work, all of these studies measure \lya\ fluxes from either narrow-band imaging or integral field spectroscopy and are therefore not subject to aperture losses from the spectroscopic slit.  A variety of recent results have shown that galaxies at both high and low redshifts are surrounded by diffuse halos of \lya\ emission \citep{sbs+11,myh+12,fhc+13,hosv+13,mon+14, mon+16}.  For UV-selected galaxies comparable to those in the current sample, \citet{sbs+11} find that spectroscopic \lya\ fluxes are $\sim5$ times lower than fluxes measured from narrow-band images on average, and $\sim3$ times lower for LAEs. For a sample of fainter LAEs, \citet{tssr15} find a smaller factor of $\sim2$ correction. 

We have no way to measure the fraction of \lya\ photons which are not captured by the slit for the extreme galaxies, but we can estimate the appropriate correction by considering other galaxy samples. The UV-continuum-selected LAEs of \citet{sbs+11} are comparable to the extreme sample in continuum luminosity and strength of \lya\ emission, while the LAEs studied by \citet{tssr15} occupy the extreme region in the BPT diagram (Trainor et al., in prep); these results suggest that a factor of 2--3 correction is likely to be appropriate for the extreme sample. 

We can also estimate the size of this correction using the results of \citet{mon+16}, who measure extended \lya\ halos around a large sample of LAEs at $z=2.2$. They find that the total \lya\ luminosity ranges from $\sim2$--4 times the luminosity within an aperture 1\arcsec\ in radius, and that the fraction in the extended halo depends on both \lya\ luminosity and equivalent width, with lower luminosity galaxies with lower equivalent widths having a higher fraction of flux in the halo. Our extreme sample spans nearly the full range in \lya\ luminosity and equivalent width considered by \citet{mon+16}, but comparing our median luminosity and equivalent widths with their results suggest that the factor of 2--3 correction determined above is reasonable.  Adjusting our escape fractions by this factor then results in a mean value of 0.16--0.24, somewhat lower than or roughly comparable to other samples of LAEs. 

The COS aperture used to measure the \lya\ emission of the Green Peas and the 1.2\arcsec\ slit used at \ztwo\ span similar physical sizes of 8--10 kpc, much larger than the typical sizes of both samples; both the Green Peas and UV-selected \ztwo\ galaxies with strong \lya\ emission have rest-frame UV sizes $r \lesssim 2$ kpc \citep{hsme15,lss+12b}. However, when aperture corrections are not applied the \lya\ escape fractions of the Green Peas are higher than those of the extreme sample, with mean (median) $f_{\rm esc}^{\lya} = 0.24$ (0.18) compared to the extreme mean of 0.08. This difference may in part be accounted for by higher extinction at \lya\ in the higher redshift sample, given the well-studied correlation between $E(B-V)$ and \lya\ equivalent width (see Section \ref{sec:extinct} above). The \ztwo\ galaxies may also be subject to larger aperture losses; while only three of the Green Peas have larger aperture \lya\ measurements (either from \textit{GALEX} or from the \lya\ Reference Sample, \citealt{ohd+14,hod+14}) which can be compared with the COS spectra, the COS aperture captures 40--75\% of the total flux for these three objects, somewhat larger than the $\sim30$--50\% we have estimated for the extreme sample. 

While the extreme galaxies have higher aperture-corrected \lya\ escape fractions than more typical star-forming galaxies at the same redshifts, the typical escape fraction remains relatively low; as in all other galaxy samples studied, the \lya\ photons of these relatively low metallicity objects are subject to significant scattering and absorption.

\section{Summary and Discussion}
\label{sec:summary}

Motivated by previous observations of strong \lya\ emission in low metallicity galaxies at \ztwo\ \citep{eps+10,stark+14}, we study the \lya\ properties of a sample of 14 galaxies lying in the upper left corner of the \OIII\ $\lambda$5007/\Hb\ vs.\ \NII\ $\lambda$6584/\Ha\ diagnostic diagram, at the low metallicity/high ionization end of the star-forming sequence (see Figure \ref{fig:bpt}). This cut is effectively a metallicity selection, as we use the O3N2 diagnostic to determine oxygen abundances, and the galaxies in this extreme sample have typical metallicities of $12+\log (\rm{O/H}) \sim 8.0$. The extreme galaxies have median stellar mass $4.0 \times 10^9$  \msun, and median dust-corrected \Ha\ star formation rate 27.8 \msunyr. Their median rest-frame UV absolute magnitude is $M_{\rm UV} = -20.66$, comparable to $M^*$ at \ztwo\ ($-20.70$, \citealt{rs09}).

The mean (median) rest-frame \lya\ equivalent width of this extreme galaxy sample is 39.3 (35.6) \AA, and 11 of the 14 objects (79\%) are \lya-emitters (LAEs) with \wlya~$> 20$ \AA. This is a much higher LAE fraction than that of a comparable sample of 522 star-forming galaxies at $2.0<z<2.6$ selected without regard to their nebular line ratios. This comparison sample has mean (median) \lya\ equivalent width $-1.0$ ($-4.4$) \AA, and 9\% of the comparison galaxies qualify as LAEs. Although the \lya\ equivalent width distribution of the extreme galaxies is quite different than that of more typical star-forming galaxies at \ztwo, it is similar to that of the local Green Pea galaxies, $z\sim0.2$ galaxies from the Sloan Digital Sky Survey selected to have very high equivalent width \OIII\ emission. The \lya\ profiles are shown in Figure \ref{fig:lyaprofiles}, and the equivalent width distributions of the extreme galaxies and comparison samples are shown in Figure \ref{fig:lyahists}.

The galaxies in the extreme sample show a range in nebular extinction as estimated from the \Ha/\Hb\ ratio (see Figure \ref{fig:bd_o3o2}). The median attenuation at \lya\ is 1.9 times lower than that of the KBSS parent sample, and 4.3 times lower if only the objects with the most robust measurements of the Balmer decrement are considered. This difference in extinction undoubtedly aids the escape of \lya\ photons in the extreme galaxies, but the range in \Ha/\Hb\ ratios suggests that moderate extinction is not a barrier to strong \lya\ emission. This result is not surprising, as some LAEs are known to be dusty \citep{kse+10,hcg+14}, although it may be more unexpected for low metallicity objects with line ratios similar to the nearly dust-free Green Peas to show significant extinction.  Future higher S/N measurements of the Balmer decrement will determine whether or not this is indeed the case. 

The median oxygen abundance of the extreme galaxies is $\sim50$\% lower than that of the parent sample, and this may also facilitate the escape of \lya\ photons, as the harder ionizing spectrum and higher ionization parameter associated with strong, low metallicity star formation may reduce the covering fraction or column density of neutral hydrogen (e.g.\ \citealt{nos+13,son+14,est+14,hsme15,tssr15}). We will explore this question further with a detailed analysis of the interstellar absorption lines and \lya\ profile structure of the extreme galaxies with high S/N rest-frame UV spectra (Erb et al.\ in prep).

The high fraction of LAEs in the extreme galaxy sample indicates that the use of nebular emission line ratios can be an effective technique for identifying samples of galaxies which are likely to have strong \lya\ emission. This method may be particularly useful at redshifts at which \lya\ is not otherwise accessible; in particular, future observations of nebular emission line ratios at $z\gtrsim6$ with {\it JWST} may allow the identification of galaxies that are likely to have a sufficiently low column density or covering fraction of neutral hydrogen to permit the escape of \lya\ photons (and, potentially, Lyman continuum radiation as well), even if the opacity of the intergalactic medium prevents observation of the \lya\ emission directly. 

We expect that studies of larger samples of galaxies with extreme line ratios at a range of redshifts will clarify the relationships between the physical conditions in \HII\ regions and the properties that ease the escape of \lya\ and LyC radiation; such an improved understanding will be required in order to fully explain the sources of the photons which reionized the universe.

\acknowledgements

We would like to thank the anonymous referee for a helpful and constructive report. DKE is supported by the US National Science Foundation through the Faculty Early Career Development (CAREER) Program, grant AST-1255591. Additional support comes from the NSF through grants AST-0908805 (CCS) and AST-1313472 (CCS, ALS), and an NSF Graduate Student Research Fellowship (ALS).  We are grateful to Alaina Henry for providing flux measurements for the Green Pea galaxies, and to the organizers and participants of the First Carnegie Symposium in Honor of Leonard Searle, ``Understanding Nebular Emission in High-Redshift Galaxies," held at the Carnegie Observatories in July 2015, for many useful discussions. Finally, we wish to extend thanks to those of Hawaiian ancestry on whose sacred mountain we are privileged to be guests.

\bibliographystyle{apj}

\begin{turnpage}
\begin{deluxetable*}{l l l c c c r r c r r r}
\tablewidth{0pt}
\tabletypesize{\footnotesize}
\tablecaption{Rest-Frame Optical Line Ratios\label{tab:neblines}}
\tablehead{
\colhead{Object} &
\colhead{RA} &
\colhead{Dec} &
\colhead{$z_{\rm neb}$} &
\colhead{$M_{\ast}$} &
\colhead{SFR$_{\rm H\alpha}$} &
\colhead{$F(\rm H\alpha)$} &
\colhead{Log(\NII/\Ha)} &
\colhead{Log(\OIII/\Hb)} &
\colhead{\Ha/\Hb} &
\colhead{\OIII/\OII} &
\colhead{12+log(O/H)} \\
\colhead{} &
\colhead{(J2000)} &
\colhead{(J2000)} &
\colhead{} &
\colhead{($10^9$ \msun)} &
\colhead{(\msun\ yr$^{-1}$)} &
\colhead{($\times 10^{-17}$)} &
\colhead{\phantom{1}} &
\colhead{\phantom{1}} &
\colhead{} &
\colhead{} &
\colhead{(O3N2)} \\ 
\colhead{(1)} &
\colhead{(2)} &
\colhead{(3)} &
\colhead{(4)} &
\colhead{(5)} &
\colhead{(6)} &
\colhead{(7)} &
\colhead{(8)} &
\colhead{(9)} &
\colhead{(10)}
}
\startdata
Q0100-BX239 & 01:03:13.709 & \phantom{-}13:15:16.128 & 2.3034 & \phantom{0}2.8 & 28.3 & $5.3\pm0.2$ & $<-1.11$ & $0.81^{+0.05}_{-0.05}$ & $4.76\pm0.60$ & $4.06\pm0.39$ & $<8.12$ \\
Q0142-BX165 & 01:45:16.867 & -09:46:03.468 & 2.3577 & \phantom{0}2.2 & 70.2 & $21.5\pm0.3$ & $-1.53^{+0.17}_{-0.12}$ & $0.80^{+0.01}_{-0.01}$ & $3.70\pm0.09$ & $3.88\pm0.13$ & $7.98^{+0.04}_{-0.05}$ \\
Q0207-BX74 & 02:09:43.154 & -00:05:50.208 & 2.1889 & \phantom{0}5.2 & 58.1 & $16.7\pm0.2$ & $-1.35^{+0.07}_{-0.06}$ & $0.94^{+0.05}_{-0.04}$ & $4.13\pm0.42$ & $8.77\pm0.34$ & $8.00^{+0.02}_{-0.03}$ \\
Q0207-BX87 & 02:09:44.234 & -00:04:13.512 & 2.1924 & 18.6 & 22.8 & $10.0\pm0.2$ & $<-1.16$ & $0.82^{+0.05}_{-0.05}$ & $3.41\pm0.38$ & $4.78\pm0.33$ & $<8.10$ \\
Q0207-BX144 & 02:09:49.210 & -00:05:31.668 & 2.1682 & \phantom{0}2.8 & 48.6 & $15.5\pm0.2$ & $<-1.44$ & $0.78^{+0.03}_{-0.03}$ & $3.98\pm0.23$ & $6.00\pm0.21$ & $<8.02$ \\
Q0821-BX102 & 08:21:05.858 & \phantom{-}31:09:53.892 & 2.4151 & \phantom{0}6.3 & 27.0 & $12.3\pm0.1$ & $<-1.60$ & $0.82^{+0.01}_{-0.01}$ & $3.01\pm0.08$ & $8.99\pm0.89$ & $<7.96$ \\
Q0821-BX221 & 08:21:06.898 & \phantom{-}31:08:08.160 & 2.3957 & \phantom{0}5.4 & 13.4 & $6.1\pm0.2$ & $<-1.21$ & $0.77^{+0.01}_{-0.01}$ & $3.03\pm0.12$ & $5.57\pm0.53$ & $<8.10$ \\
Q1217-BX164 & 12:19:41.100 & \phantom{-}49:40:11.028 & 2.3310 & \phantom{0}2.2 & 83.3 & $5.7\pm0.2$ & $-1.14^{+0.16}_{-0.11}$ & $0.79^{+0.11}_{-0.09}$ & $7.38\pm1.58$ & $2.40\pm0.17$ & $8.11^{+0.05}_{-0.06}$ \\
Q1700-BX553 & 17:01:02.201 & \phantom{-}64:10:39.396 & 2.4719 & \phantom{0}2.8 & 21.8 & $8.3\pm0.2$ & $<-1.10$ & $0.75^{+0.03}_{-0.03}$ & $3.18\pm0.22$ & $7.16\pm0.45$ & $<8.14$ \\
Q1700-BX711 & 17:01:21.288 & \phantom{-}64:12:20.664 & 2.2947 & 20.0 & 19.0 & $10.7\pm0.2$ & $-1.25^{+0.10}_{-0.08}$ & $0.80^{+0.02}_{-0.02}$ & $2.79\pm0.12$ & $4.89\pm0.59$ & $8.08^{+0.03}_{-0.03}$ \\
Q2206-BM64 & 22:08:52.354 & -19:43:28.416 & 2.1943 & \phantom{0}8.5 & 27.4 & $7.8\pm0.1$ & $-1.21^{+0.17}_{-0.12}$ & $0.75^{+0.05}_{-0.05}$ & $4.15\pm0.49$ & $4.08\pm0.26$ & $8.10^{+0.04}_{-0.06}$ \\
Q2343-BX418 & 23:46:18.571 & \phantom{-}12:47:47.364 & 2.3054 & \phantom{0}2.8 & 52.0 & $14.0\pm0.1$ & $-1.34^{+0.08}_{-0.07}$ & $0.87^{+0.02}_{-0.02}$ & $4.02\pm0.20$ & $6.83\pm0.31$ & $8.02^{+0.02}_{-0.03}$ \\
Q2343-BX460 & 23:46:17.134 & \phantom{-}12:48:06.912 & 2.3945 & \phantom{0}1.9 & 14.1 & $5.8\pm0.1$ & $-1.38^{+0.16}_{-0.11}$ & $0.79^{+0.03}_{-0.02}$ & $3.19\pm0.19$ & $4.38\pm0.30$ & $8.04^{+0.04}_{-0.05}$ \\
Q2343-BX660 & 23:46:29.434 & \phantom{-}12:49:45.552 & 2.1742 & \phantom{0}5.4 & 28.8 & $18.5\pm0.2$ & $<-1.69$ & $0.79^{+0.02}_{-0.02}$ & $2.52\pm0.13$ & $12.15\pm0.35$ & $<7.93$
\enddata
\tablecomments{Column descriptions: (1) Galaxy ID (2--3) RA and Dec (J2000) (4) Systemic redshift from joint fit to 
the nebular emission lines. (5) Stellar mass in units of $10^9$ \msun, assuming a 
\citet{c03} initial mass 
function. (6) Star formation rate from extinction-corrected \Ha\ flux using the relationship 
of \citet{k98}. (7) \Ha\ flux in units of $10^{-17}$ erg s$^{-1}$ cm$^{-2}$.
(8) Ratio of \NII\ $\lambda6584$/\Ha. (9) Ratio of \OIII\ $\lambda5007$/\Hb.
(10) Observed \Ha/\Hb\ ratio with statistical errors; see text for discussion of additional systematic uncertainties due to slit loss corrections.
 (11) Observed ratio \OIII\ $\lambda\lambda 4959, 5007$/\OII\ 
$\lambda\lambda 3727, 3729$, without correcting for extinction. (12) Oxygen abundance 
calculated with the O3N2 calibration of \citet{pp04}.}
\end{deluxetable*}
\end{turnpage}

\begin{deluxetable*}{l c c r r r r r}
\tablewidth{0pt}
\tabletypesize{\footnotesize}
\tablecaption{\lya\ Properties\label{tab:lya}}
\tablehead{
\colhead{Object} &
\colhead{$z_{\rm neb}$} &
\colhead{$z_{\rm Ly\alpha}$} &
\colhead{$F(\rm Ly\alpha)$} &
\colhead{$W_{\rm Ly\alpha}$} &
\colhead{$\Delta v_{\rm Ly\alpha}$} &
\colhead{\lya/\Ha} &
\colhead{$f_{\rm esc}^{\rm Ly\alpha}$} \\
\colhead{} &
\colhead{} &
\colhead{} &
\colhead{($\times 10^{-17}$)} &
\colhead{(\AA)} &
\colhead{(\kms)} &
\colhead{} &
\colhead{}  \\
\colhead{(1)} &
\colhead{(2)} &
\colhead{(3)} &
\colhead{(4)} &
\colhead{(5)} &
\colhead{(6)} &
\colhead{(7)} &
\colhead{(8)}
}
\startdata
Q0100-BX239 & 2.3034 & 2.3067 & \phantom{1}$5.3\pm0.4$ & $18.3\pm2.0$ & $296\pm 54$ & $1.01\pm0.09$ & $0.04\pm0.01$ \\
Q0142-BX165 & 2.3577 & 2.3590 & $28.2\pm0.4$ & $61.2\pm2.3$ & $114\pm 15$ & $1.31\pm0.03$ & $0.09\pm0.01$ \\
Q0207-BX74 & 2.1889 & 2.1882 & $27.3\pm0.7$ & $101.9\pm4.4$ & $-61\pm 36$ & $1.64\pm0.04$ & $0.09\pm0.02$ \\
Q0207-BX87 & 2.1924 & 2.1930 & $33.2\pm0.7$ & $77.3\pm4.7$ & $ 54\pm 21$ & $3.33\pm0.09$ & $0.27\pm0.07$ \\
Q0207-BX144 & 2.1682 & 2.1695 & $19.8\pm0.9$ & $37.5\pm2.0$ & $125\pm 61$ & $1.28\pm0.06$ & $0.07\pm0.01$ \\
Q0821-BX102 & 2.4151 & ... & $<0.78$ & $<-3.25$ & ... & $<0.06$ & $<0.007$ \\
Q0821-BX221 & 2.3957 & 2.3987 & \phantom{1}$0.9\pm0.3$ & $7.7\pm2.5$ & $269\pm212$ & $0.15\pm0.05$ & $0.02\pm0.01$ \\
Q1217-BX164 & 2.3310 & 2.3340 & \phantom{1}$2.9\pm0.1$ & $22.0\pm2.2$ & $270\pm 19$ & $0.51\pm0.03$ & $0.007\pm0.003$ \\
Q1700-BX553 & 2.4719 & 2.4747 & \phantom{1}$9.7\pm0.5$ & $33.7\pm2.4$ & $240\pm 43$ & $1.16\pm0.07$ & $0.11\pm0.02$ \\
Q1700-BX711 & 2.2947 & 2.2958 & \phantom{1}$2.2\pm0.5$ & $20.4\pm5.3$ & $100\pm212$ & $0.21\pm0.05$ & $0.02\pm0.01$ \\
Q2206-BM64 & 2.1943 & 2.1973 & \phantom{1}$7.9\pm1.0$ & $22.9\pm3.4$ & $281\pm 85$ & $1.02\pm0.12$ & $0.05\pm0.02$ \\
Q2343-BX418 & 2.3054 & 2.3077 & $32.8\pm1.2$ & $61.8\pm2.6$ & $208\pm 57$ & $2.34\pm0.09$ & $0.13\pm0.02$ \\
Q2343-BX460 & 2.3945 & 2.3966 & \phantom{1}$7.8\pm0.4$ & $47.7\pm2.8$ & $184\pm 71$ & $1.35\pm0.07$ & $0.13\pm0.02$ \\
Q2343-BX660 & 2.1742 & 2.1778 & $16.8\pm0.7$ & $40.4\pm2.2$ & $338\pm 43$ & $0.91\pm0.04$ & $0.10\pm0.02$
\enddata
\tablecomments{Column descriptions: (1) Galaxy ID (2) Systemic redshift from joint fit to the nebular emission lines. 
(3) \lya\ redshift from centroid of \lya\ emission. (4) \lya\ flux in units of $10^{-17}$ erg s$^{-1}$
  cm$^{-2}$. (5) Rest frame \lya\ equivalent width. (6) Velocity offset of the centroid of \lya\ emission 
  from the systemic redshift. (7) Observed \lya/\Ha\ flux ratio. (8) \lya\ escape fraction, assuming an 
  intrinsic ratio \lya/\Ha~$=8.7$ and with \Ha\ fluxes corrected for extinction based on the \Ha/\Hb\ ratio.}
\end{deluxetable*}

\end{document}